\documentclass{elsart}
\usepackage{epsfig}
\usepackage{amssymb}

\begin{document}
\begin{frontmatter}

\title{Accidental Politicians: How Randomly Selected Legislators can Improve Parliament Efficiency} 

\author
{Alessandro Pluchino} \ead{alessandro.pluchino@ct.infn.it}
\address{Dipartimento di Fisica e Astronomia, Universit\`a di
Catania, and INFN sezione di Catania, - Via S. Sofia 64, I-95123
Catania, Italy} 

\author
{Cesare Garofalo}  \ead{cesaregarofalo@yahoo.com}
\address {Dipartimento di Analisi dei Processi Politici, Sociali e Istituzionali, Universit\'a di Catania, Via Vittorio Emanuele II 8, I-95131 Catania, Italy}

\author
{Andrea Rapisarda} \ead{andrea.rapisarda@ct.infn.it}
\address{Dipartimento di Fisica e Astronomia, Universit\`a di
Catania, and INFN sezione di Catania, - Via S. Sofia 64, I-95123
Catania, Italy} 

\author
{Salvatore Spagano}  \ead{salvo.spagano@gmail.com}
\address {Dipartimento di Economia e Metodi Quantitativi, Universit\'a di Catania,
Corso Italia 55, I-95100 Catania, Italy}

\author
{Maurizio Caserta}  \ead{caserta@unict.it}
\address {Dipartimento di Economia e Metodi Quantitativi, Universit\'a di Catania,
Corso Italia 55, I-95100 Catania, Italy}

\begin{abstract}
We study a prototypical model of a Parliament with two Parties or two Political Coalitions and we show how the introduction of a variable percentage of randomly selected independent legislators can increase the global efficiency of a Legislature, in terms of both the number of laws passed and the average social welfare obtained. We also analytically find an "efficiency golden rule" which allows to fix the optimal number of legislators to be selected at random after that regular elections have established the relative proportion of the two Parties or Coalitions. These results are in line with both the ancient Greek democratic system and the recent discovery that the adoption of random strategies can improve the efficiency of hierarchical organizations. 
\end{abstract}

\end{frontmatter}

\section{Introduction}

In ancient Greece, the cradle of democracy, governing bodies were largely selected by lot \cite{Aristotle,Headlam,Sinclair}. The aim of this device was to avoid typical degenerations of any representative institution \cite{Michels}. In modern democracies, however, the standard is choosing representatives by vote through the Party system. Debate over efficiency of Parliament has therefore been centred on voting systems, on their impact on parliamentary performances and, ultimately, on the efficiency of economic system \cite{Buchanan1,Olson,Downs,Cooter}. In recent years also physicists have started to provide a quantitative understanding of social and economical phenomena \cite{Helbing,Schweitzer,Epstein2,Klimek,caruso1,social-atom,Fortunato} and it is in this perspective that the present  work should  be placed. 
In this paper, rediscovering the old Greek wisdom and recalling a famous diagram about human nature by C.M.Cipolla \cite{Cipolla}, we show how the injection of a measure of randomness improves the efficiency of a parliamentary institution. In particular, we present numerical simulations of the efficiency of a prototypical Parliament modeled by means of  an agent based model \cite{Wilensky}. We also  find an analytical expression, whose predictions are confirmed by the simulations, that determines the exact number of randomly selected legislators, in an otherwise elected parliament, required to optimize its aggregate performance. The latter is estimated by the number of approved acts times the average social gain. This result, on  one hand is in line with the positive role which random noise plays often in nature and in particular in physical systems \cite{caruso2,frenkel,spagnolo}.   
On the other hand, it goes also in the same direction of the recent discovery \cite{Pluchino1,Pluchino2} that, under certain conditions, the adoption of random promotion strategies improves the efficiency of human hierarchical organizations in order to face the problem of the so-called "Peter Principle" \cite{Peter}.
\\
The paper is organized as follows. In Section 2 we describe the Parliament model and its dynamics. In Section 3 we present the main numerical and analytical results. Then, in Section 4, we discuss several historical examples in order to give an empirical support to our findings. Finally, conclusions and remarks are drawn.

\section{The Parliament Model}

Human societies need institutions \cite{Coase1,North}, since they set the context for individuals to trade among themselves. They are expected, therefore, to have an impact on the final outcome of those trading relations \cite{Coase3}. This paper looks at a specific institution, the Parliament, designed to hold the legislative power and to fix the fundamental rules of society. 

%
\begin{figure}  
\begin{center}
\epsfig{figure=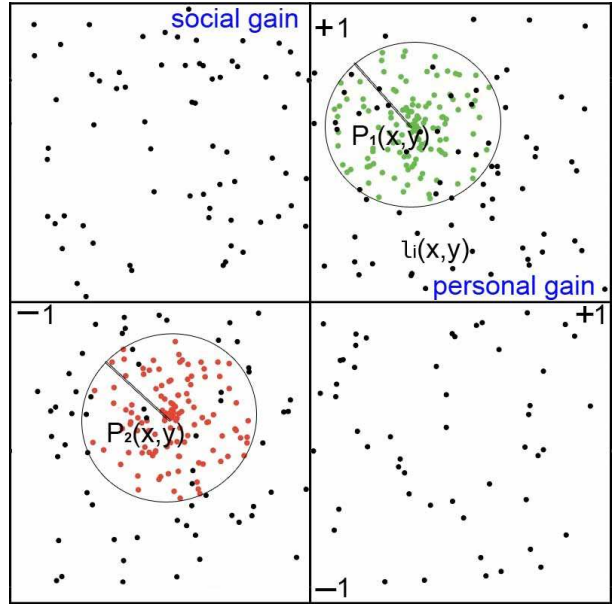,width=8truecm,angle=0}
\end{center}
\caption{{\it  Cipolla Diagram. Each point in this diagram, with coordinates in the intervals $[-1,1]$, represents a member of Parliament, according to his/her attitude to promote personal or social interests. The Parliament consists of $N=500$ members: black points represent $N_{ind}=250$ independent legislators, while green and red points refer to the remaining members, belonging to the two Parties $P_1$ and $P_2$. We report also the {\it circles of tolerance} of the two Parties, with equal radius r=0.3, see text for further details. Please notice that some free points could apparently fall within the circle of tolerance of some Party, but of course the correspondent legislators will remain independent. }}
\label{Fig.1}
\end{figure}

\subsection{The Cipolla Diagram}

A Parliament can be modeled as resulting from the aggregate behavior of a number of legislators, who are expected to make proposals and vote. In so doing they are pictured as moved by personal interests, like re-election or other benefits, and by a general interest. Taking both motivations into account, it is possible to represent individual legislators as points $l_i(x,y)$ (with $i=1,...,N$) in a diagram (see Fig.1), where we fix arbitrarily the range of both axes in the interval $[-1,1]$, with personal gain on the $x$-axis and social gain (understood as the final outcome of trading relations produced by law) on the $y$-axis. Each legislator will be therefore described through his/her attitude to promote personal and general interest. 
\\
This diagram takes after a very famous one proposed in 1976 by the economic historian Carlo M. Cipolla  \cite{Cipolla},  who represented human population according to its ability to promote personal or social interests. Of course people do not always act consistently, therefore each point in the Cipolla diagram represents the weighted average position of the actions of the correspondent person. 
\\
The basic idea of this study is to use the Cipolla classification in order to elaborate a prototypical agent based model \cite{Wilensky} of a Parliament with only one Chamber, consisting of $N=500$ members and $K=2$ Parties or Coalitions, and to evaluate its efficiency in terms of both approved acts and average social gain ensured.
In particular, all the points $l_{jk}(x,y)$ representing members of given Party $P_k$ will lie inside a circle with a given radius $r_k$ and with a center $P_k(x,y)$ falling in one of the four quadrants.
The center of each Party is fixed by the average collective behavior of all its members, while the size of the respective circle indicates the extent to which the Party tolerates dissent within it: the larger the radius, the greater the degree of tolerance within the Party. Therefore, we call the circle associated to each Party {\it circle of tolerance}.
\\
It is clear that, in real Parliaments, the fact of belonging to a Party increases, for a legislator, the likelihood that his/her proposals are approved. But it is also quite likely that the social gain resulting from a set of approved proposals will be on average reduced if all the legislators fall within the influence of some Party (more or less authoritarian). In fact, even proposals with little contribution to social welfare will be approved if Party discipline prevails, while, if legislators were allowed to act according to their judgement, bad proposals would not receive a large approval.
Therefore, the main goal of this paper is to explore how the global efficiency of a Parliament may be affected by the introduction of a given number $N_{ind}$ of independent members, i.e. randomly selected legislators free from the influence of any Party, which will be represented as free points on the Cipolla diagram. The independent members, once randomly selected for a given legislature, should not be candidates in any successive legislature, to avoid the risk of being "captured" by existing Parties or Coalitions.

\subsection{Dynamics of the Model}

The dynamics of the model is the following. During a Legislature $L$ each legislator (agent) $l_i(x,y)$ (independent) or $l_{jk}(x,y)$ (belonging to Party $P_k$) can perform only two simple actions: (i) proposing an act and (ii) voting (for or against) a proposal. 
\\
The first action does not depend on the membership of the agent: each legislator proposes one or more acts of Parliament ($a_n$, with $n = 1,...,N_a$, being $N_a$ the total number of acts proposed by all the legislators during the Legislature $L$), with a given personal and social advantage depending on his/her position on the diagram (i.e. $a_n(x,y) \equiv l_i(x,y)$ for every act proposed). It follows that legislators belonging to a Party can propose acts which are not perfectly in agreement with the Party's common position, as function of their distance from the center $P_k(x,y)$ of the correspondent circle of tolerance.
\\
The action of voting for, or against, a proposal is more complex and strictly depends on the membership of the voter and on his/her {\it acceptance window}. The acceptance window is a rectangular window on the Cipolla diagram into which a proposed act $a_n(x,y)$ has to fall in order to be accepted by the voter, whose position fixes the lower left corner of the window (see Fig.2). This follows from the assumption that we imagine ideal legislators who are able to recognize better or worse proposals than their ones, but only accept proposals better than (or equal to) their ones. 
%
\begin{figure}  
\begin{center}
\epsfig{figure=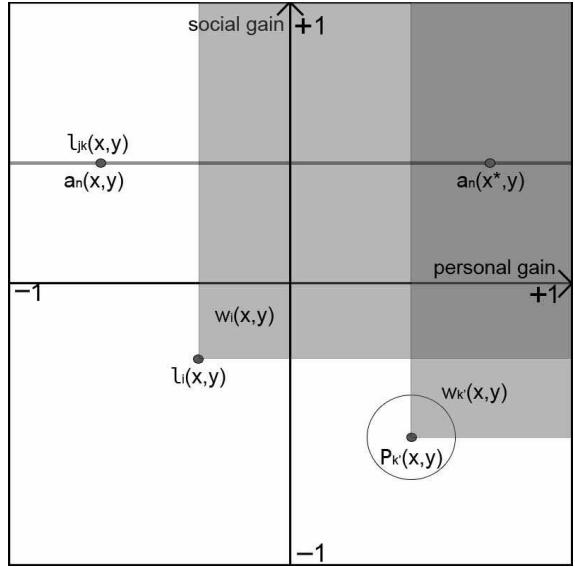,width=7.5truecm,angle=0}
\end{center}
\caption{ {\it In this Figure we show an example of the voting process. The two grey areas in Cipolla Diagram indicate the acceptance windows of an independent legislator $l_i(x,y)$ and of a Party $P_{k'}$. Given the proposal $a_{n}(x,y)$, advanced by a legislator $l_{jk}(x,y)$ belonging to the Party $P_{k}$, a new "voting point" $a_{n}(x^*,y)$ has been randomly extracted over the (gray) line $y=y(a_n)$ and compared with the two acceptance windows. Since the voting point falls within the window $w_{i}(x,y)$ of $l_i$, this legislator will vote the correspondent act. The same proposal would be voted also by all the members $l_{jk'}$ of the Party $P_{k'}$, since it also falls within the window $w_{k'}(x,y)$ (see text).}}  
\label{Fig.2}
\end{figure}
The main point is that, while each free legislator has his/her own acceptance window, so that his/her vote is independent from the othersÕ vote, all the legislators belonging to a Party always vote by using {\it the same} acceptance window, whose lower left corner corresponds to the center of the circle of tolerance of their Party. Furthermore, following the Party discipline, any member of a Party accepts {\it all} the proposals coming from any another member of the same Party. 
\\
It is now important to stress that, while the perception of the social advantage $y(a_n)$ (i.e. the $y$-coordinate) of a given act $a_n$ can be likely considered as unambiguously determined for each legislator or Party, the perception of the personal advantage $x(a_n)$ (i.e. the $x$-coordinate of $a_n$) cannot. Indeed, the fact that a certain $a_n(x,y)$ would be favorable for a given legislator, does not imply that it should be favorable for another legislator or for a Party. Therefore, the coordinate $x(a_n)$ of any proposed act has to be different for any legislator or Party and will be expressed by a random number $x^*$, uniformly extracted in the interval $[-1,1]$: it is this new position $a_n(x^*,y)$, called {\it voting point} and lying on the line $y=y(a_n)$ (see Fig.2), that has to be compared with the acceptance windows of legislators and Parties. It follows that, from  the personal advantage point of view, the act can be either approved or rejected depending on whether the voting point lies on the right or left of the acceptance window corner.
Finally, once {\it all} the $N$ members of Parliament voted for or against a certain proposal, the latter will be accepted only if receives at least ${N\over2}+1$ favorable votes. We indicate with $N_{acc}$ the overall number of {\it accepted} proposals.   
At this point we need some global quantity which in some way would be able to express the efficiency of the Parliament during a Legislature $L$. 
\\
An immediate measure of the Parliament activity could be the percentage of accepted acts over the total $N_a$, i.e. $N_{\%acc}(L)=(N_{acc}/N_a)*100$. But another important quantity is surely the average social welfare ensured by all the accepted acts of Parliament, expressed by $Y(L)={N_{acc}}^{-1} \sum_{m=1}^{N_{acc}}  y(a_{m})$.
It is therefore convenient to take the product of these two quantities in order to obtain the {\it global efficiency} of a Legislature: 

\begin{equation}
\label{efficiency}
Eff(L)=N_{\%acc}(L)*Y(L).
\end{equation}

that is expressed by a real number included in the interval [-100,100]. We anticipate that, in order to obtain a measure independent of the particular configuration of Parliament, the global efficiency (\ref{efficiency}) has to be further averaged over many Legislatures, each one with the same number of proposals but with a different distribution of legislators and Parties on the Cipolla diagram. 

\section{Simulation Results}

In our simulations we will focus on a Parliament with $N=500$ members distributed over two Parties or Coalitions $P_1$ and $P_2$ with different relative sizes, a configuration which is simple and interesting at the same time, being typical of many Countries with a bipolar political system. We will also consider a set of $N_L=100$ Legislatures, each one with a total number of $N_a=1000$ proposals. In order to study how the global efficiency of the Parliament depends on the number $N_{ind}$ of independent legislators, let us consider, first, the two limiting cases $N_{ind}=0$ and $N_{ind}=N$. 

\subsection{Parliament with 2 Parties and $N_{ind}=0$}

The efficiency of a Parliament without independent legislators strictly depends, for a given Legislature $L$, on the random position of the centers of the two Parties, with coordinates, respectively, $x(P_1)$, $y(P_1)$ and $x(P_2)$, $y(P_2)$ over the Cipolla diagram, but also on their size (in terms of percentage of members) and on the radius $r$ of their circle of tolerance. Suppose to assign a percentage $p=60\%$ of legislators to $P_1$ and the remaining ($40\%$) to $P_2$. Let us consider a sequence of $N_a=1000$ acts of Parliament, proposed each time by a randomly chosen legislator. Actually, in the further limiting case of a radius $r = 0$ (i.e. all the positions of the members of a Party coincide with its center), we have the following two possibilities, each one with probability $1/2$: 
\\
(i) If $y(P_2)<y(P_1)$, only the acts of Parliament coming from Party $P_1$ will be accepted, since members of $P_1$ will never vote for any proposal coming from $P_2$, and the percentage of accepted proposal during the Legislature $L$ will be equal to the percentage $p$ of members of Party $P_1$, i.e. $N_{\%acc}(L)\sim60\%$; it follows that the average social welfare $Y(L)$ of the accepted acts should approximately coincide with the $y$-coordinate of $P_1$, i.e. $Y(L) \sim y(P_1)$.
\\
(ii) If $y(P_2)>y(P_1)$, in addition to all the proposals of $P_1$, will be also accepted those proposals of $P_2$ which will randomly fall in the acceptance window of $P_1$. This will depend on the coordinate $x(P_1)$ and will occur with a probability $1-x(P_1) \over 2$. Such a probability is $1$ for $x(P_1)=-1$, $0$ for $x(P_1)=1$ and, on average, is equal to $1/2$. Therefore, being the number of proposals coming from $P_2$ approximately the $40\%$ of the total, 
in average only $20\%$ of those should be accepted, thus yielding with respect to the previous case an increase in both the percentage of accepted proposals and the average social welfare. Of course non-null values of the radius $r$ will produce slight modifications in these predictions, since the positions of the members of a Party start to spread within the circle of tolerance.
\\
\begin{figure}  
\begin{center}
\epsfig{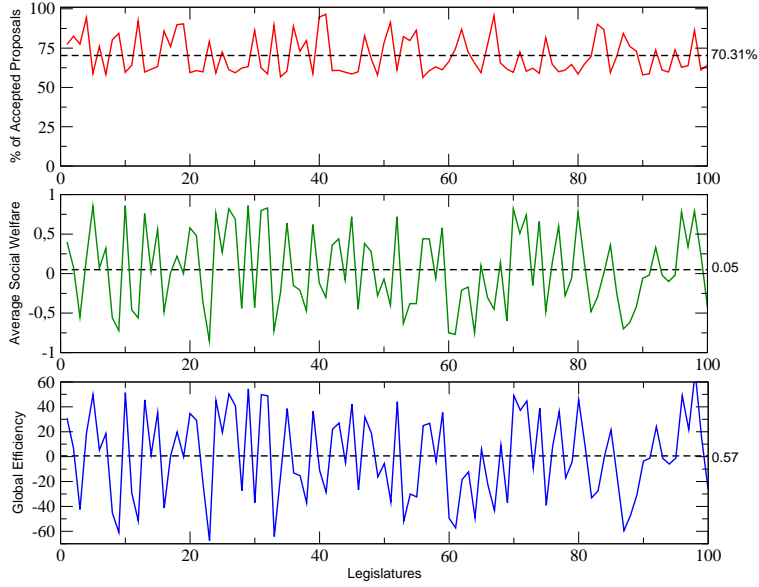}
\end{center}
\caption{{\it Simulation results for a Parliament with $500$ members distributed into two Parties $P_1$ and $P_2$ with, respectively, the $60\%$ and $40\%$ of legislators (as shown in Fig.2) and radius $r_1=r_2=r=0.1$. No independent legislators are present ($N_{ind}=0$). In the three panels we plot (from top to bottom), for a set of $N_L=100$ Legislatures (each one with a different position of $P_1$ and $P_2$ over the Cipolla diagram), the percentage of accepted proposals, the average social Welfare and the correspondent global Efficiency, calculated as the product of the previous two quantities. For each Legislature $L$ an array of $N_a=1000$ proposals has been considered. The averaged values for the three quantities are reported on the right and are sketched with a dashed line inside the panels.}}
\label{Fig.3}
\end{figure}
In Fig.3 we show the simulation results obtained for the set of $N_L=100$ Legislatures, each one with a different position of $P_1$ and $P_2$ over the Cipolla diagram. A small radius $r=0.1$, equal for both the Parties, has been chosen in order to check the predictions obtained for $r=0$. For each Legislature $L_h$ ($h=1,...,N_L$) the correspondent values of $N_{\%acc}(L_h)$, $Y(L_h)$ and $Eff(L_h)$ have been plotted in three distinct panels (from top to bottom). At the end of the simulation the average values $AV(N_{\%acc})$,  $AV(Y)$ and $AV(Eff)$ have been calculated and reported in the panels as dashed lines (with the respective numerical value on the right side). 
\\
As expected, the number of accepted proposals $N_{\%acc}(L_h)$ oscillates, with probability $1/2$, between $60\%$ (case (i)) and a number around $80\%$ (case (ii)), thus producing an $AV(N_{\%acc})\sim 70\%$, while the value $Y(L_h)$ oscillates between $-1$ and $1$ thus producing an almost null average value $AV(Y)=0.05$ (this is the case because the values $y(P_1)$ will result uniformly distributed along the $y$-axis when considers the entire set of $100$ Legislatures). Consequently, also the product $N_{\%acc}(L)*Y(L)$ will oscillate around zero and the average efficiency of the Parliament $AV(Eff)=0.57$ is quite small (notice that, following the definition, the range of variation of $Eff(L_h)$ is $[-100,100]$). This means that a Parliament without legislators free from the influence of Parties turns out to be rather inefficient (as probably happens in reality).   

\begin{figure}  
\begin{center}
\epsfig{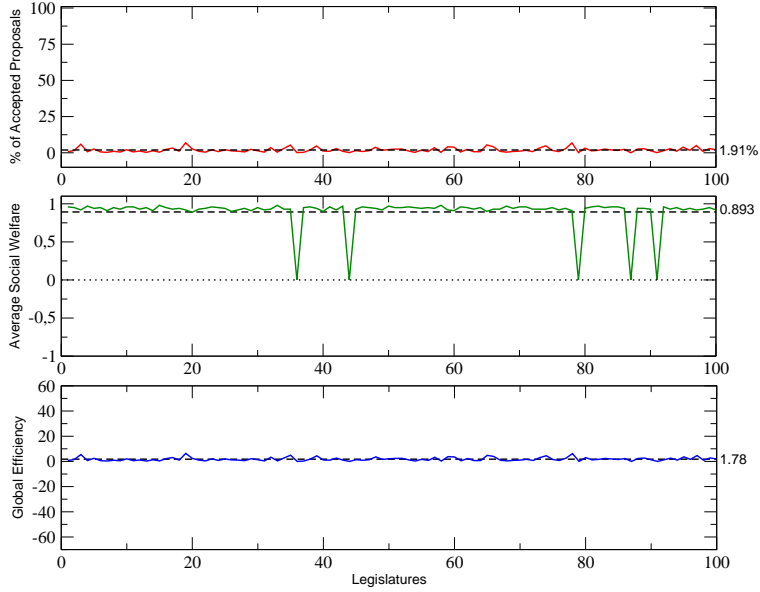}
\end{center}
\caption{{\it Simulation results for a Parliament with ($N_{ind}=500$) independent members, without any Parties. As in the previous figure, in the three panels we plot (from top to bottom), for a set of $N_L=100$ Legislatures (each one with a different distribution of free points, representing the independent legislators), the percentage of accepted proposals, the average social Welfare and the correspondent global Efficiency. For each Legislature $L$ an array of $N_a=1000$ proposals has been considered. The averaged values for the three quantities are reported on the right and are sketched with a dashed line inside the panels.   }}
\label{Fig.4}
\end{figure}

\subsection{Parliament with no Parties and $N_{ind}=N$}

Let us consider, now, the opposite situation in which only independent legislators make up in the Parliament. In this case no Parties exist and the points $l_i(x,y)$, corresponding to the $N=500$ members of Parliament, are uniformly distributed over the Cipolla diagram. 
It is evident that now a given act of Parliament $a_n(x,y)$ will be accepted only if the majority ${N \over 2}+1$ of these points will fulfill the prescriptions $y(l_i)< y(a_n)$ and $x(l_i)< x^*(a_n)$, being $x^*(a_n)$ the $x$-coordinate of the voting point $a_n(x^*,y)$, randomly extracted with uniform distribution over the straight line of equation $y=y(a_n)$ for each legislator $l_i$ which is requested to vote for $a_n$. 
\\
Being $l_i(x,y)$ uniformly distributed on the plane, for a given value of $y(a_n)$ only about $50\%$ of the $\tilde{N}(a_n)$ legislators with $y(l_i)<y(a_n)$ will accept the proposal. 
But such a number will be clearly lower than ${N \over 2}$ unless $y(a_n)\sim1$, since only in this latter case the half-plane $y<y(a_n)$ will coincide with the entire Cipolla diagram and ${\tilde{N}(a_n)\over2} \sim {N\over2}$. Thus it follows that, during a Legislature $L$, only a very small number of proposals will be accepted, but with a very high social gain $y(a_n)\sim1$. 
\\
In Fig.4 we show analogous simulations as in Fig.3, but in this opposite case with $100\%$ of independent legislators ($N_{ind}=500$). It results that, averaging again over $100$ Legislatures, the previous predictions are confirmed: as expected, we find a very small value for $AV(N_{\%acc}) \sim 2\%$ (top panel) and a very high value for the social Welfare $AV(Y)=0.893$ (middle panel: here, the few points with $Y(L)=0$ correspond to Legislatures with $N_{\%acc}=0\%$). But it follows that, as in the case with $0\%$ independent legislators, the efficiency of a Parliament with only independent members will be quite small again: as it is visible in the bottom panel of Fig.4, the product $N_{\%acc}(L)*Y(L)$ will stay near zero for all the Legislatures, thus giving an average global efficiency $AV(Eff)=1.78$. It looks therefore that no particular benefit stems from abolishing Parties altogether.
\\
Now, once having explored these two limiting cases, it is interesting to see how the efficiency of a Parliament with two Parties (or, more in general, two political coalitions) is affected by the increase of the number of independent members from $N_{ind}=0$ to $N_{ind}=500$.

%
\begin{figure}  
\begin{center}
\epsfig{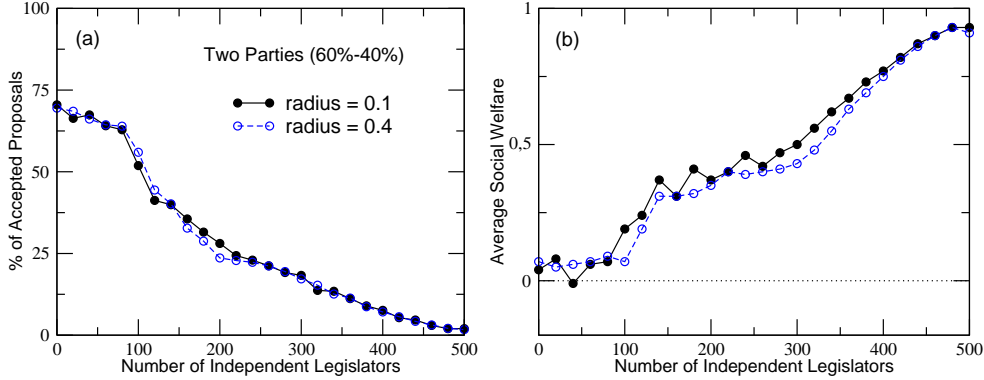}
\end{center}
\caption{{\it Simulation results for a Parliament with $N = 500$ members, two Parties $P_1$ and $P_2$, with circles of tolerance of two sizes, i.e. $r_1=r_2=0.1$ (full circles) and $r_1=r_2=0.4$ (empty circles), and an increasing number of independent legislators $N_{ind}$. For each value of $N_{ind}$, we distribute the remaining legislators ($N-N_{ind}$) into the two Parties with percentages $60\%$ and $40\%$.  
In Panel (a) we show the average number of accepted proposals, $AV(N_{\%acc})$, which monotonically decreases with $N_{ind}$. In Panel (b), on the contrary, the average value of the social Welfare $AV(Y)$ is shown to increase with $N_{ind}$. See text for further details.}}
\label{Fig.5}
\end{figure}

\subsection{Configurations with 2 Parties and an increasing number of independent legislators} 
 
In the simulations shown in Fig.5  we vary from $0$ to $N$ the number $N_{ind}$ of independent legislators in a Parliament with $N=500$ members and we distribute the remaining legislators ($N-N_{ind}$) into the two Parties with percentages $60\%$ and $40\%$ and with radius $0.1$ and $0.4$ (equal for the two Parties). We see that, increasing $N_{ind}$, (i) the average number of accepted proposals $AV(N_{\%acc})$ decreases from $\sim70\%$ to $\sim2\%$, Panel (a), while, on the other hand, (ii) the average value of the social Welfare $AV(Y)$ increases from $\sim0$ to $\sim0.9$, Panel (b). In both cases the increase/decrease is monotonic and it is not much influenced by the value of $r$. At this point it is interesting to explore what is the behavior of the {\it product} of these two quantities, i.e the behavior of the global efficiency $AV(Eff)$ of the Parliament, emerging from the interplay between the accepted proposals and the social welfare they provide. Actually, we observed that $AV(Eff)$ turns out to be quite small in both the limiting cases with $0$ and $N$ independent legislators (see Figs.3 and 4, bottom panels), so it makes sense to ask what happens in the intermediate region $0<N_{ind}<N$. 
%
\begin{figure}  
\begin{center}
\epsfig{figure=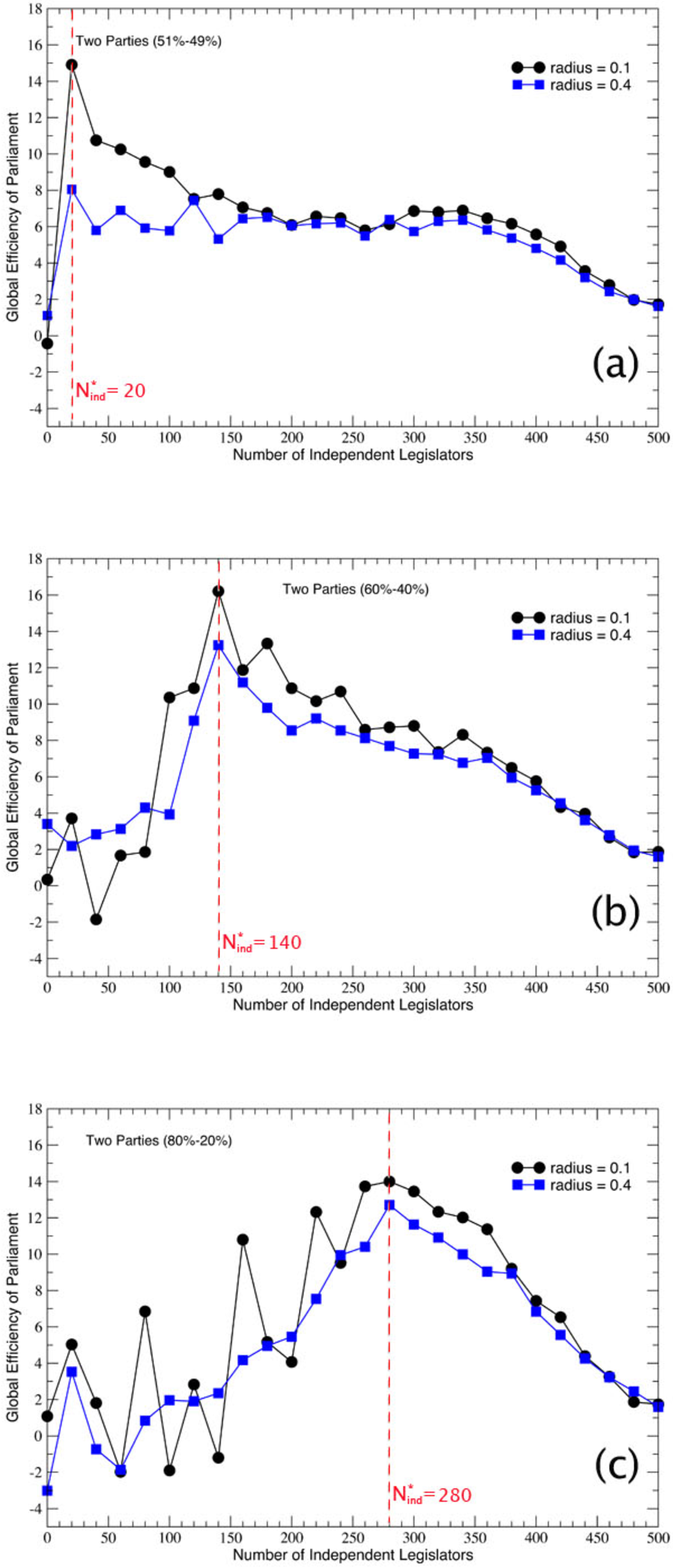,width=8truecm,angle=0}
\end{center}
\caption{{\it The global efficiency of a Parliament with $N = 500$ members and two Parties $P_1$ and $P_2$, with circles of tolerance of two sizes, $r=0.1$ (circles) and $r=0.4$ (squares), is plotted as function of an increasing number of independent legislators $N_{ind}$. Each point represents an average over $100$ Legislatures, each one with $1000$ proposals of acts of Parliament coming from randomly selected legislators. The three panels differ in the percentage of the $(N-N_{ind})$ members assigned to the two Parties. In Panel (a): 51\%, 49\%; Panel (b): 60\%, 40\%; Panel (c): 80\%, 20\%. In all the panels, for a specific $N_{ind}^*$ (indicated by a vertical dashed line), it is visible a peak in efficiency which shifts from left to right going from the top to the bottom panel (see text).}}
\label{Fig.5}
\end{figure}
\\
In Fig.6 we answer to this question by plotting the global efficiency of the Parliament as function of an increasing number of independent legislators. In particular, the case $60\%-40\%$, that we considered in Fig.3 and Fig.4, is plotted in panel (b). The two curves corresponds to two different sizes of the circles of tolerance, with radius, respectively, $r=0.1$ and $r=0.4$ (equal for both the Parties). Each point of a given curve represents, as usual, an average over $100$ Legislatures, each one with $1000$ proposals of acts of Parliament. The result is very interesting, because it clearly appears that the efficiency, albeit with a fluctuating behavior, rapidly increases with the number of independent legislators until it reaches a peaked maximum for $N_{ind}^*=140$, then smoothly decreases towards the known limiting value $\sim2$. This means, on  one hand, that the introduction of any number of independent legislators, out of the influence of any Party, in general improves the efficiency of a Parliament and, on the other hand, that exists an optimal percentage of these legislators which makes the efficiency of the Parliament the highest possible. It is also interesting to observe that the maximum value of efficiency decreases when the radius $r$ increases from $0.1$ to $0.4$, implying that the constructive role of independent legislators is sensitive to the degree of freedom in the Parties: the more authoritarian Parties are (cases $r=0.1$), the more the role of independent legislators becomes decisive. When, on the contrary, Parties are libertarian (cases $r=0.4$), efficiency tends  to no longer depend on the number of independent members.   
\\
We also simulated different configurations of the Parliament, changing the size of the two Parties. In panel (a) of Fig.6 we assigned percentages of $51\%$ and $49\%$ to the Parties, and the result is that the peak in efficiency occurs much early than before, with $N_{ind}^*=20$ independent legislators. Such an effect is quite reasonable because it suggests that, if the two competing Parties have a similar size, even a small number of independent members working in the Parliament, playing a role of balance, can fairly improve its global efficiency. On the other hand, when one Party is quite bigger that the other one, like in the simulations shown in the panel (b), the number of independent legislators required to enhance the efficiency of the Parliament increases. 
\\
This trend is confirmed in panel (c) of Fig.6, where a Parliament with one Party very much bigger ($80\%$) than the second one ($20\%$) has been considered: as expected, after a very oscillating initial behavior, the (smoother) peak shifts on the right with respect to the previous panels and the maximum efficiency is obtained with the introduction of $N_{ind}^*=280$ independent legislators. Finally, also in the last two panels the efficiency depends on the radius of the circles of tolerance of the two Parties, and in general decreases as it is increased. Notice that in all the cases, while the values of  $AV(Eff)$ for  $N_{ind}=0$ can be different for the curves with different radius, going towards $N_{ind}=N$ the two curves tend to coincide, since the dynamics (and therefore the efficiency) becomes independent of the radius of the Parties.

\subsection{The Efficiency Golden Rule}
 
Unlike the relatively simple behavior of the system in the two limiting cases $N_{ind}=0$ and $N_{ind}=N$, the general case with $0<N_{ind}<N$ is much more complex to manage and  it is absolutely not trivial to predict the efficiency value of the peaks shown in the three panels of Fig.6, which stays approximately constant although it depends on all the features that affect the voting process. However, quite surprisingly, a simple formulation exists to work out the optimal number $N_{ind}^*$ of independent legislators as function of the size $p$ (in percentage) of the majority Party.
\\
Actually, we could argue that, in a given Legislature with two Parties of different sizes, none of which holding the absolute majority of the members in the Parliament (due to the presence of independent members), $N_{ind}^*$ would be in some way associated to the minimum number of independent legislators which, added to the majority Party $P_1$, allows it to reach the threshold of  ${N \over 2}+1$ members necessary to accept a given proposal $a_n$. But we know from the previous subsection that, being the independent legislators $l_j(x,y)$ uniformly distributed over the Cipolla diagram, for a given value of $y(a_n)$ only about $50\%$ of all the independent members with $y(l_j) < y(a_n)$ will vote for the proposal, equivalent to the $\tilde{N}(a_n)$ points lying in the left half of the half-plane below the line of equation $y=y(a_n)$. Such a rule continues to apply also when Parties exist. In this case, being a generic $y(a_n)$ randomly distributed over the $y$-axis throughout many Legislatures, we can safely say that, on average, the line $y=y(a_n)$ will coincide with the $x$-axis, therefore, for a given $N_{ind}$, only $N_{ind} \over 4$ independent members (i.e. those lying in the left half of the half-plane $y<0$) will vote the proposal. Thus, in order to find $N_{ind}^*$, we just need to add this number to the number of members of the majority Party $P_1$, i.e. $(N-N_{ind}^*) \cdot {p \over 100}$, and to impose the following equality:

\begin{equation}
\label{equality}
(N-N_{ind}^*)\cdot\frac{p}{100} + \frac{N_{ind}^*}{4} = \frac{N}{2} + 1
\end{equation}

Finally, solving this equation with respect to $N_{ind}^*$, one easily obtains:
 
\begin{equation}
\label{goldenrule}
N_{ind}^*= \frac{2N - 4N\cdot(p/100) + 4}{1 - 4\cdot(p/100)}.
\end{equation}
 
This prediction closely matches the numerical results of simulations performed for several values of $p$, as shown in Fig.7. We checked that it is true for several sizes $N$ of the Parliament and that is independent of the number of Legislatures and of the number of proposals for each Legislature. Being also independent of the radius of the circles of tolerance of the Parties, we argue that Eq.(\ref{goldenrule}) could be considered an universal {\it golden rule} for optimizing the efficiency of any social situation with two competing groups of elected people through the introduction of randomly selected independent voters. 
 %
\begin{figure}  
\begin{center}
\epsfig{figure=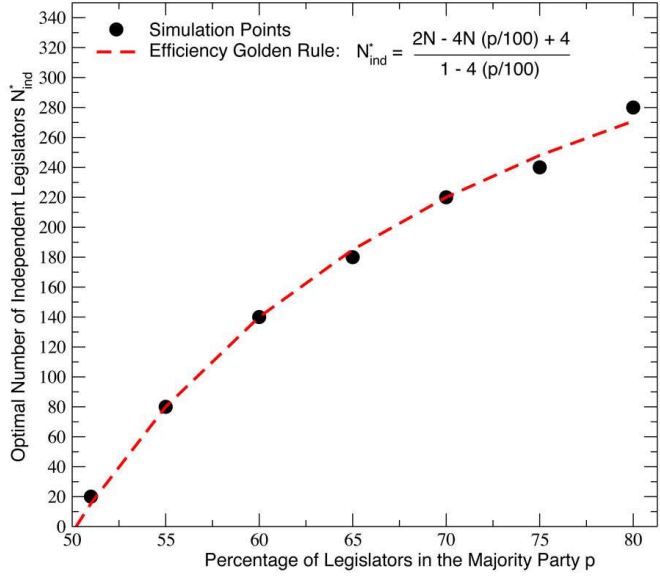,width=10truecm,angle=0}
\end{center}
\caption{{\it The optimal number of independent legislators $N_{ind}^*$ is plotted (full circles) as function of the size $p$ (in percentage) of the majority Party $P_1$, for our Parliament with $N = 500$ members and two Parties. An average over $100$ Legislatures, each one with $1000$ proposals of acts of Parliament, has been performed for each point. This plot is invariant for values of the radius of the Parties in the range $[0.1,0.5]$. The dashed line represents the prediction of the efficiency Ôgolden ruleÕ, reported also in the upper part of the figure (see text).}}
\label{Fig.3}
\end{figure}
\\
The reason why Eq.(3) performs quite well can be easily grasped with the aid of intuition: since any majority, held by a Party larger than 50 $\%$, brings no contribution to social welfare (being $AV(Eff)\sim0$ for $N_{ind}=0$), one could easily go without it. However, bringing the majority down to 50 $\%$ will allow for a larger share of independent legislators. Such legislators, on  one hand (in a measure equal to $N_{ind}/4$) will help the relative majority Party to retain its majority, on the other (again in a measure equal to $N_{ind}/4$) will offer proposals with a positive contribution to social welfare which are bound to be voted by that very majority. So nothing is lost from the positive role of the Party system (effectiveness in voting) and all is gained from the role of independent legislators (the quality of their proposals). Thinking of a practical application for a real Parliament, the knowledge of the golden rule would allow to fix the optimal number of accidental politicians to be chosen at random, by picking them up from a given list of candidates (i.e. ordinary citizens fitting the requirements), after that regular elections have established the relative proportion of the two Parties or Coalitions. 
\\
As a last remark, we also observe that from Eq.(3) it follows immediately that one has $N_{ind}^* > {N\over2}$ for ${p\over100} > {3\over 4} + {2\over N} $ (with $N \ge 8$). This means that, when $N>>2$, the optimal number $N_{ind}^*$ of independent legislators overcomes the threshold for reaching the majority in the Parliament if the biggest Party exceeds $p_{th}=75\%$ (notice that, for $N \rightarrow 8$, ${p\over100} = {3\over 4} + {2\over N} $ tends to $1$ and the threshold value $p_{th}$ tends to $100\%$). Actually, we saw that in the case $80\%-20\%$ (panel (c) of Fig.6) our Parliament with $N=500$ ($>>2$) members reached its maximum efficiency with $N_{ind}^*=280$ independent legislators: therefore, in this case the independent legislators alone would have the majority of the seats. This is another reason for imposing the constraint that, in general, randomly selected legislators should not be candidates in any successive legislature, to reduce also the risk that they could join and vote together acting as a new majority Party.

\section{Discussion and Historical Review}

In this section we discuss some historical  facts which provide  substantial and successful empirical  support to  our results.
In fact, for a modern political observer, our findings could probably sound very strange.
Today, most people think that democracy means elections, i.e. believe that only electoral mechanism could ensure representativeness in democracy. However, as already mentioned in the introduction, in the first significant democratic experience, namely the Athenian democracy, elections worked side by side with random selection ({\it sortition}) and direct participation. Actually, in that period Parties did not exist at all and random selection was the basic criterion when the task was impossible to be carried out collectively in the Assembly, where usually Athenian citizens directly made the most important decisions. Of course only the names of those who wished to be considered were inserted into the lottery machines, the {\it kleroteria} \cite{Aristotle,Headlam,Sinclair}.
\\  
Sortition was not used in Athens only. Probably, already others Greek city-states adopted the Athenian method, even if historical documentation is dubious. For sure, many other cities in the history used some kind of lot as rule, such as Bologna, Parma, Vicenza, San Marino, Barcelona and some parts of Switzerland. Lot was also used in Florence in the 13th and 14th century and in Venice from 1268 until the fall of the Venetian Republic in 1797, providing opportunities to minorities and resistance to corruption \cite{Mowbray}. 
\\
In the course of history, little by little, the concept of representativeness overlapped with that of democracy, until it became its synonymous. Consequently, today, in contemporary institutions, almost any random ingredient has been expunged. Among the few historical vestiges of sortition, there are the formation of juries in some judicial process and the selection by lot in some public policy \cite{Boyle}.
Actually, even if nowadays the information and communication technology would revitalize the possibility of direct democracy, (the so-called {\it E-democracy}), this idea meets opposition as much as random selection, since the representative system, and his correlated Party system, is strongly believed to be the only way to make society a democratic place.  
\\
On the other hand, the drawbacks of Party system have been well documented. For example, the Òiron law of oligarchyÓ of the sociologist Robert Michels  \cite{Michels}, states that all forms of organization, democratic or not, inevitably develop into oligarchies. The indispensability of leadership, the tendency of all groups to defend their interests and the passivity of represented people, are only a few of the many reasons that deteriorate every democratic Party system. In the representative democracy this process is even institutionalized. Party elites act to serve the Party and themselves, often at the expenses of the public interest. If some members of Parliament vote against the Party line on any issue, these are likely to be ostracized, expelled, or not endorsed at the next elections. 
\\
Of course free elections are an indubitable progress in comparison with authoritarian regimes, but today the electoral system tends to form a Òdemocratic aristocracyÓ, where representatives are superior to the electorate. In particular, the representativeness is organized by ideas, or classes, which flow together into the Parties' programs. Unfortunately, this kind of organizations (in turn, institutions) very slowly accept social changes because of various rents that they ensure. 
Therefore, in the last decades, the idea of choosing representatives by random selection has been re-introduced in political thinking and gained a fair number of supporters \cite{Carson,Amar,Mulga,Elster,Barnett}. Demarchy, or statistical democracy, is the name proposed by someone \cite{Martin}. According to these supporters there would be several advantages in the sortition method. 
For example, social and demographic features (income, race, religion, sex,) would get a fair distribution in the parliament, so the interest of the people would get a more effective representativeness and politically active groups in society, who tend to be those who join political Parties, would not be over-represented. On the other hand, representatives appointed by sortition do not owe anything to anyone for their position, so they would be loyal only to their conscience, not to political Party, also because they are not concerned in their re-election. Furthermore, sortition may be less corruptible than elections. It is easy to ensure a totally fair procedure by lot. On the contrary the process of elections by vote can be subject to manipulation by money and other powerful means. 
\\
In this context our results, on  one hand, provide a quantitative  confirmation of  the poor efficiency of a Parliament based only on Parties or Coalitions (see the case $N_{ind}=0$) and, on the other hand, sustain in a more  rigorous way  the constructive role of independent, randomly selected, legislators.  
In any case, it is worthwhile to stress that, in our model, the electoral system is not abolished altogether, but only integrated with a given (exactly determined) percentage of randomness, a feature which could be implemented in a simple way in real  systems. 
\\
Before closing this section, let us remark that our findings are also perfectly in line with recent studies \cite{Pluchino1,Pluchino2} about the effectiveness of random promotion strategies in hierarchical organizations. In these studies, the  strategy of promoting people at random has been shown to be, under certain conditions, a successful way to circumvent the effects of the so-called "Peter Principle" \cite{Peter} and to increase the efficiency of a given pyramidal or modular organization. We think that random selection could be of help in contrasting Peter Principle effects also in the context of parliamentary institutions, which are exposed to analogous risks linked to the change of competences required to the elected people in their new political positions.

\section{Conclusions and Remarks}

In this paper, by means of a prototypical Parliament model based on Cipolla classification, we showed in a quantitative way  that the  introduction of  a well-defined number of random members into the Parliament improves the efficiency of this institution through the maximization of  the social overall welfare that depends on its acts. In this respect, the exact number of random members has to be established {\it after} the elections, on the basis of the electoral results and of our analytical "golden rule": the greater the size difference between the Parties, the greater the number of members that should be lotted to increase the efficiency of Parliament \cite{suppl-material}. 
\\
Of course our prototypical model of Parliament does not represent all the real parliamentary institutions around the world in their detailed variety, so there could be many possible way to 
extend it. 
For example it would be interesting to study the consequences of different electoral systems by introducing more than two Parties in the Parliament, with all the consequences deriving from it. 
Also the government form could be important: our simple model is directly compatible with a presidential system, where there is no relationship between Parliament and Government, whereas, in the case of a parliamentary system, also such a link should to be considered in order to evaluate the overall social welfare. For simplicity, we chose to study a unicameral Parliament, whereas several countries adopt bicameralism. So, simulating another chamber could bring to subsequent interesting extensions of the model. Finally, we expect that there would be also several other social situations, beyond the Parliament, where the introduction of random members could be of help in improving the efficiency.
\\
In conclusion, our study provides  rigorous arguments in favor of the idea that the introduction of random selection systems, rediscovering the wisdom and the history of ancient democracies, would be broadly beneficial for modern institutions.

\end{document}